\documentstyle[12pt]{article}
\begin{document}
\title{COLOR TRANSPARENCY ASSUMPTIONS}
\author{D.~Makovoz and G.~A.~Miller}
\date{November 1, 1994}
\maketitle
{\em Department of
Physics, FM-15, University of Washington, Seattle, WA 98195,   USA}\\
\vspace{1.0cm}\\
\begin{abstract}
An exactly solvable model is used to investigate the assumptions behind color
transparency.
\end{abstract}
\vspace{1.0cm}
\section{ INTRODUCTION}

Color transparency (CT) is the
 anomalously high  transparency of the nucleus to nucleons in
quasi-elastic high-momentum transfer nuclear processes, measured with
resolution good enough to insure
that no ``extra" pions are produced.
This means that
the absorptive nuclear optical potential representing
initial and final state interactions plays no role in such reactions.
Color transparency is under active investigation with experiments performed at
BNL~\cite{bnl88}, SLAC~\cite{ne18}
 and ongoing work at BNL~\cite{eva} and experiments proposed at CEBAF.
Further references can be found in the review Ref.~\cite{fms94}.

The existence of color transparency depends upon three
assumptions~\cite{Mue82,Bro82}:

(i) A small wave packet is formed in a high momentum transfer reaction.
This wave packet is sometimes dubbed a point like configuration (PLC).

(ii) The interactions between a small color neutral
wave packet and the nucleus are suppressed.

(iii) The wave packet escapes the nucleus before expanding.
The expansion time $\tau$  is typically stated as $\tau
\sim\tau_0 P/M$
where $\tau_0$ is a time in the rest frame (expected to be about
1 fm) needed for a small system to
expand and $P/M$ is a time dilation factor. For large enough momentum, P,
$\tau$ is large.

Each of these assumptions can be questioned. Indeed, the question of whether
or not a small wave packet is formed depends crucially on properties of the
ground state wavefunction, hence on poorly understood features of
non-perturbative  quantum chromodynamics QCD~\cite{FMS92,FMS931}.
The reduction of the final state interaction is often explained as due to
the cancellation of gluon emission amplitudes that occurs when a
color singlet system consists of closely separated quarks and gluons.
The legality of adding amplitudes (before squaring) requires  a
coherent reaction, so that the cancellation is limited to a select class
of  reactions. An additional issue is the value of the factor $M$ in
the time dilation; why couldn't it be as large  as
$~P$? Another way to phrase this
question is that
the expansion of the PLC can be slow only if
highly excited intermediate states
are not important \cite{jm90,jm91}.

The purpose of this paper is to use a simple
model with realistic features
to investigate the meaning and limitations
of each of the three basic assumptions.
The model is defined and described in Sec. II. Each ``nucleon" is the
ground state of an electrically neutral systems of two quarks
interacting via the Coulomb potential.
The wave functions for this potential
can not be computed using pertrubation theory, but can
be computed exactly. In this sense
the above force provides an example of a non-perturbative, yet
solvable interaction.
The two quarks
have different masses, with the ratio of the heavier to the lighter being two.
The nucleus is a collection of such nucleons. This model is relevant
to color transparency  physics because of the
analogy between the electric neutrality of quantum electrodynamics
QED and the color neutrality of
QCD. Within this model,
it is straightforward to construct (Sect. III) the
wavepacket formed in a high momentum transfer reaction.
We take the hard interaction as simply adding momentum to the heavy quark.
This wave packet evolves as it moves through the
surrounding nucleus. This evolution is governed  by the internal Coulomb force
between the heavy and light  quarks and also the final state Coulomb
interaction with the surrounding nucleons.
This interaction is computed, and
the dependence
on the transverse separation between the heavy and light quarks is
presented
in Sec. IV.
The process we consider in this paper is meant to
be analogous to the (e,e'p) scattering, one of the candidate reactions for the
observation of CT.
In the (e,e'p) reaction hard and soft interactions are of different
nature - the first one is electromagnetic, the second one is strong. The
smallness of the soft interaction is a consequence of the color neutrality
of the proton, which does also have electric charge.
In our model
the soft interaction is of
electromagnetic nature,
 and  our ``nucleon" is electrically neutral. The high momentum
transfer interaction has a
distinct charge associated with it (only one of the quarks has this
charge).
The numerical results
of the calculations are given in Sec. V.
The properties of this model are that the size of
the wave packet created in a high momentum $Q$ transfer process declines with
$Q$, and  that the
 expansion rate of this wave packet is inversely proportional to
$Q$.
The origin of this slow expansion rate is investigated in Sec.~VI.
A brief summary is presented in Sec. VII.

\section{THE MODEL}

We consider a nonrelativistic, electrically neutral system of two quarks
interacting with the Coulomb potential.
The strength of the the electromagnetic interaction,
is given by the fine structure constant $\alpha$;
$\alpha$ determines the ratio of  the binding energy  of the system
 to the mass of its constituents. In Nature $\alpha=1/137$, but
within the present model $\alpha$  can be varied
   and need not be small.  Thus the interaction we
use is intended to represent the strong color force.

The form of the bound state and continuum wave functions are well
known,
and are not presented here. It is worthwhile to discuss the system of
units that we use.
Coulomb units with  $\hbar=1$, $e=1$, and
the mass of the light quark $m_{q}=1.5$ (to set the reduced mass $m_{red}$ to
1)
 are  the natural choice for this problem.
 With this choice the size of the system in the ground state, the Bohr
radius $r_{B}$, is the unit of length,
 and twice the ground state energy is the unit of energy.
 Note that in the Coulomb units
 it is  the speed of light $c=1/\alpha$ that sets the energy scale.
 In this paper  $1/\alpha$ will be used instead of $c$.
A relativistic dispersion relation is used for the energy $E_{X}$
of the state $|X>$ with momentum $P_{X}$
\begin{equation}
E_{X}^{2}=M_{X}^{2}/\alpha^{4}+ P_{X}^{2}/\alpha^{2} .  \label{eq:dispersion}
\end{equation}
 The rest mass $M_{X}$ includes the binding energy
$\varepsilon_{X}$ as well as the quark masses
\begin{equation}
M_{X}=4.5+\varepsilon_{X}\alpha^2 \label{eq:restmass}.
\end{equation}
 The energy eigenvalues  are
\begin{equation}
\varepsilon_{n}=-1/2n^{2}
\end{equation}
for the discrete part of the spectrum and
\begin{equation}
\varepsilon_{k}=k^{2}/2.
\end{equation}
 for the continuous part of the spectrum.

\section{EXPANDING WAVE PACKET}
A wavepacket, possibly a point-like configuration,
is created when a photon of momentum $Q$ is
absorbed by the heavy quark of the ``nucleon" in the ground state $|1>$.
Thus we write
\begin{equation}
|PLC>\equiv e^{i\vec q\cdot {1\over 3}\vec r}|1>,\label{eq:hard}
\end{equation}
where spin is ignored, ${1\over 3} \vec r$ is the position operator for the
heavy quark relative to the center of mass. Thus Eq.~(\ref{eq:hard})
defines our model of the hard high momentum transfer reaction.
 The transferred momentum is defined as
$\vec q =Q \hat z$, which specifies the direction of the z-axis.

The use of completeness allows us to express
this wavepacket in terms of the elastic $F(Q^{2})$ and
inelastic  $f_{X}(Q^{2})$ form factors
\begin{equation}
|PLC>= F(Q^{2})|1>+ \sum_{X}(\int dX) f_{X}(Q^{2})|X> \label{eq:PLC}
\end{equation}
and
\begin{equation}
F(Q^{2})=\int d^{3}r \psi^{*}_{1}(r) \exp (i\vec{q} \cdot {1\over 3}\vec{r})
\psi_{1}(r),
\end{equation}
\begin{equation}
f_{X}(Q^{2})=
\int d^{3}r \psi^{*}_{1}(r) \exp (i\vec{q} \cdot {1\over 3}\vec{r})
\psi_{X}(r).
\end{equation}
where $\psi_{X}(r)$ are the Coulomb eigenfunctions (subscript 1 for
the ground state).
 The summation (integration) is over
the complete set of the Coulomb eigenstates, $X\equiv nl$ for the
discrete states and $X\equiv kl$ for the continuum states.

If there were no final state interaction (FSI) of the hadronic wavepacket
with the nucleus the amplitude of detecting the ``nucleon" in the ground
state would be equal to the form factor $F(Q^{2})$.
Color transparency is concerned with situations in which experimental
kinematics constrain the final
state interactions to be soft, of low momentum transfer. We shall model these
soft interactions by treating the
nuclear medium here as a set of neutral
``nucleons", Sect.~IV. Then
the soft interactions are  approximately proportional to the
product of the wave-packet-``nucleon"
 forward scattering
amplitude, with the density of nucleons.
In the impulse approximation, the interaction  is expressed in terms
of
a matrix element of an operator $\hat{\chi}(b)$.
The main feature of this operator is its dependence on $b$,
the transverse separation of the quarks $ (\vec b \cdot \hat z\equiv 0)$.
In particular, $\hat{\chi}(b=0)=0.$  We shall discuss a specific model
for
$\hat{\chi}$ in Sect. IV below.

With this notation, the
scattering amplitude ${\cal M}_1$ is given by
\begin{equation}
{\cal M}_1(Z)=<1|\hat{\chi}\hat{G}(Z)|PLC>, \label{eq:calm}
\end{equation}
to first order in $\hat{\chi}$ where
$\hat{G}(Z)$ is the Green's propagator
of a  PLC a distance Z (along the $\hat q $ direction) through the
medium. Color transparency occurs if this
term is small compared to the Born
amplitude $F(Q^2)$. Note that we use lower case letters (b,z) to denote
transverse and longitudinal
quark-antiquark separations and upper case letters (B,Z) to denote the
displacement of the center of the wave packet from the center of the
nucleus. An evaluation of the matrix element of Eq.~(\ref{eq:calm})
requires an integration over $d^3r$, but not over the coordinates
(B,Z). (The B-dependence of ${\cal M}_1$  is suppressed for simplicity.)

In the standard Glauber treatment of final state interactions
\cite{glauber} an optical potential
approximation is often used. The potential is proportional to the
forward scattering amplitude, hence to the total nucleon-nucleon
cross section $\sigma$.
This cross section determines the rate of the exponential decay of the
scattering nucleon wave function. In the present case the
b-dependence of $|PLC>$ varies with Z, because the Green's function
$\hat G$ includes the effects of the
heavy quark light quark Hamiltonian.
Thus the PLC forward
scattering amplitude varies with Z. In particular, if
the initial state of Eq.~(\ref{eq:PLC}) corresponds to one of very small
transverse size, the PLC expands as Z increases.
We shall assume that the
eikonal approximation is valid. In that case, each of the states from
the complete set of states defined above in
(\ref{eq:PLC}),
acquires a phase $\exp iP_{X}Z$ as it propagates a distance $Z$.

It is useful to define an effective cross section. $\sigma_{eff}(Z)$,
with
\begin{equation}
\sigma_{eff}(Z)\equiv {\cal M}_1/F(Q^2).
\end{equation}
This quantity depends on the  overlap of $\hat \chi(b)$
with the quark-antiquark wave function, and is therefore
a measure of how the  size of the wave packet
varies with Z \cite{jm91}.
Some standard manipulations then lead to the result:
\begin{eqnarray}
\sigma_{eff}(Z)=\sigma+
\sum_{l=0,2} \sum_{n=2}^{\infty}
\frac{\chi_{nl}  f_{nl}(Q^{2})}{F(Q^{2})}
   \exp i(P_{n}-P)Z+         \nonumber\\
\sum_{l=0,2} \int_{0}^{\infty} dk
\frac{\chi_{kl}  f_{kl}(Q^{2})}{F(Q^{2})}\exp i(P_{k}-P)Z   .
           \label{eq:sigma}
\end{eqnarray}
This result is similar to the one of Jennings and Miller \cite{JM1}.
In that work the matrix elements $\chi_X$ and the inelastic form factors
$f_X$ were taken from available data and the color transparency
condition $\sigma_{eff}(Z=0)=0$ was imposed. In the present work
we use a specific model to evaluate those terms and can determine whether or
not the color transparency condition is satisfied.

To proceed, we further specify our notation.
The first term of Eq.~(\ref{eq:sigma}) is the total cross section for the
``nucleon" ground state to interact with a target ``nucleon". This is
\begin{equation}
\sigma =  <1|\hat{\chi}|1> .
\end{equation}
The matrix elements $\chi_{X}$ are
\begin{equation}
\chi_{X}=<1|\hat{\chi}|X>.
\end{equation}
The orbital angular momentum of the states $X$ are limited to even values
by the  requirements of parity conservation. We restrict the sum to values
of $l=0, 2$ to anticipate a specific form: $\hat \chi(b)\propto b^2$.

The momentum $P_{X}$ of the excited state $|X>$ is given by the
energy conservation relation imposed by the wave equation:
\begin{equation}
P_{X}^{2} +M_{X}^{2}/\alpha^2=P^{2} +M_{1}^{2} /\alpha^2,\nonumber
\\
P\simeq Q.
\end{equation}

An important quantity is
$\sigma_{eff}(Z=0)$, which  measures the size of the initially formed PLC.
This must be small for  $\sigma_{eff}(Z)$ to be small. Note that
if $\sigma_{eff}(Z=0)$ is to be small,
cancellations  in eq.~(\ref{eq:sigma})
have to render
$\sigma_{eff}(0)$ fall off rapidly with $Q$ for $Q \gg 1$.
We shall work with a simple form of the soft interaction
$\hat{\chi}(b)=\sigma\hat{b}^2/<1|\hat{b}^2|1>$; see the next section.
This form
 allows the evaluation of
$\sigma_{eff}(Z=0)$ with the result
\begin{equation}
\sigma_{eff}(0)= \frac{2}{Q^{2}r_B^2/4+1} .   \label{eq:sigma_theory}
\end{equation}
This is significant, it says that the effective size $b^2\sim 1/Q^2$ for
large $Q^2$. This is a property of the ground state
Coulomb wave function \cite{FMS92,FMS931}. PLC formation is allowed
in this model.

For non-zero values of Z
the phase factors $\exp i(P_{X}-P)Z$
spoil the cancellation of different terms in
the sum (\ref{eq:sigma}). As a result $\sigma_{eff}$ grows as Z increases from
0. This indicates that
an expanding PLC generally  experiences final state interactions.
If the PLC leaves the nucleus without significant expansion then
the final state interactions are suppressed.
How fast a particular term goes out of phase and upsets the
cancelation depends on its momentum $P_{X}$
\begin{equation}
P_{X}=\sqrt{P^2-\Delta M_X^2/\alpha^2}   \label{eq:P_X}
\end{equation}
with
\begin{equation}
\Delta M_n^2/\alpha^2= \frac{n^2-1}{n^2}(4.5-
\alpha^2\frac{n^2+1}{4n^2})
\label{eq:px}
\end{equation}
for the discrete spectrum and
\begin{equation}
\Delta M_k^2/\alpha^2= (k^2+1)(4.5+\alpha^2/4 (k^2-1))
\end{equation}
for the continuous spectrum.

Suppose that a limit $P\rightarrow \infty$ can be taken, such that all
relevant
$P_X\rightarrow P$.  In that case, the Z-dependence of $\sigma_{eff}$
disappears and
the PLC does not expand. For the discrete states
$P_n\gg \Delta M_n/\alpha$  is true for $P$ as low as
$\sim3/r_B$.
The situation is different for the continuous states, since their
energy is not bound from above.  The value of $P$ large
enough to
ensure a slow rate of the PLC expansion  depends on the structure
of the matrix elements $\chi_{kl}$ and form factors $f_{kl}$.
This goes back to the idea of Ref.~\cite{jm90},
that CT can be observed only for momenta transfer greater that
the energy of all important intermediate excited states.
We shall use specific calculations within our model to investigate these
issues. The use of equations like Eq.~(\ref{eq:sigma}) in eikonal
expressions for PLC wave functions is discussed in Ref.~\cite{greenberg}.

There is another concern about
the contribution of the higher excited states.
  The eikonal approximation used to derive the phase factors $\exp
i(P_{X}-P)Z$  breaks down if $P_X\rightarrow0$.
Therefore  any
evidence, that the contribution into the sum (\ref{eq:sigma}) of the
states with $P_X\ll 1$  is  important, invalidates the approach developed
above.

We shall also investigate the dependence on $\alpha$.
The physical range of
$\alpha$ is
from 0 to 3. 
   The value $\alpha=0 $
corresponds to the non-relativistic limit in which
the speed of light
(1/$\alpha)$ is infinite
with
\begin{equation}
E_X= \varepsilon_X + \frac{P_X^2}{2m_{red}}.
\end{equation}
The upper limit $\alpha =3$ corresponds to $M_1=0$,
recall Eq.~(\ref{eq:restmass}). A further
increase in $\alpha$ would yield a negative rest-mass of
ground state
``nucleon".

\section{THE WAVE PACKET NUCLEON      \newline
INTERACTION}

We are concerned here with deriving the interaction $\hat \chi$, which has been
defined above as the
forward scattering amplitude between two quarks (with a mass ratio of two)
of transverse
separation b and a ``nucleon"
target. We are using a non-relativistic Coulomb bound state
for the ground state dynamics of the heavy and light quarks,
so we also take the
``nucleon" targets  to be the  ground state of the same system.

The expression for the
projectile-nuclear forward scattering amplitude is \cite{glauber}
\begin{equation}
\hat f(\theta=0,b) = {ik\over 2\pi}\int d^2B\left[1-
e^{-i\chi(B,b,\hat B\cdot\hat b )}\right],
\label{eq:amp}
\end{equation}
where the integration is over the area of the nuclear target and
k is the momentum of the wave packet. The phase shift
function $\chi(B,b,\hat B\cdot\hat b)$
 is given by the integral over $dZ$ of the sum of the
light and heavy quark Coulomb  potentials.
One usually sees expressions in which the integral over $b$ times appropriate
wave functions is performed. Here we are dealing with a wave packet that is
a coherent sum of physical states, so it is convenient to study
$\hat f$.
Straightforward
manipulations lead to the result
\begin{equation}
\chi(B,b,\hat B\cdot\hat b) =
{-i\alpha 4\pi\over v}\int{d^2q_\perp\over (2\pi)^2}
{\tilde{\rho}(q_\perp)\over q^2_\perp}\;e^{i\vec q_\perp\cdot\vec B}\left(e^{i
{2\over 3}\vec q_\perp\cdot\vec b}
-e^{-i{1\over 3}\vec q_\perp\cdot\vec b}\right)
\label{eq:chi}
\end{equation}
where $v$ is the speed  of the wave packet and $\tilde{\rho}(q_\perp)$ 
 is the Fourier transform of the nucleonic charge density, $\rho(r)$:
\begin{equation}
\rho(r)=\psi_1^2({2\over 3}r)-\psi_1^2({1\over 3}r).
\end{equation}
In particular
\begin{equation}
\tilde{\rho}(q_\perp) = {1\over \left(1+{q_\perp^2r_B^2\over 9}\right)^2} -
{1\over \left(1+{q_\perp^2r_B^2\over 36}\right)^2}
\end{equation}
where the first (second) term
the Fourier transform of the charge density of the light (heavy) quark.
The neutrality condition is
\begin{equation}
\tilde{\rho}(q_\perp=0) = 0,\label{eq:0}
\end{equation}
which is vital in obtaining the result that $\chi(B,b,\hat B\cdot\hat b)$
vanishes at $b=0$.
This is because
the integral over $d^2q_\perp$ is convergent only because of Eq.~(\ref{eq:0}).
Note also that $\tilde{\rho}(q_\perp)/q_\perp^2$ is proportional to the
Fourier transform of the Coulomb potential.

We stress that
the expression Eq.~(\ref{eq:chi}) is obtained by summing coherently the
Coulomb interactions of both the heavy and light quarks with target nucleons.
This coherence, and the consequent ``small interactions at small
separations" is lost if one is computing the cross section for an inclusive
 process in which one sums over all final states of the ejected wave packet.

We first assess Eq.~(\ref{eq:amp})
by expanding the exponent in powers of
$\chi$. If the usual fine structure
constant  is used, the leading term will dominate.
There is no term of 0'th order in $\chi(B,b,\hat B\cdot\hat b) $.
The first order term
$\int d^2B\;\chi(B,b,\hat B\cdot\hat b)$ vanishes for all values of $b$
because the integration over $d^2B$ sets
$q_\perp$ to 0. 
However,
$\int d^2B\;\chi^2(B,b,\hat B\cdot\hat b)$ does not vanish.
Keeping this term leads to the result
\begin{equation}
\hat f(\theta=0,b) = {ik\over 2\pi} ({4\pi \alpha\over v})^2
\int {d^2q_\perp \over 2\pi} {\tilde{\rho}^2(q_\perp)\over q_\perp^4}
2\left(1-J_0(q_\perp b)\right).
\end{equation}
The zero'th order cylindrical Bessel function has small argument limit
$$\lim_{x\to 0} J_0(x)=1-x^2/4$$ so that
one immediately finds
\begin{equation}
\hat f(\theta = 0,b)\sim ib^2.
\end{equation}
This term is
purely imaginary, so that its influence is to exponentially damp
scattering wave functions.

We next turn to a complete evaluation of Eq. (\ref{eq:amp}).
The integration can be simplified by replacing $\vec B$ by a shifted value
$\vec B-{1\over 6}\vec b$ so that
Eq.(\ref{eq:chi}) becomes
\begin{equation}
\chi(B,b,\hat B\cdot\hat b) =
{-i\alpha 4\pi\over v}\int{d^2q_\perp\over (2\pi)^2}
{\tilde{\rho}(q_\perp)\over q^2_\perp}\;e^{i\vec q_\perp\cdot\vec B}\left(e^{i
{1\over 2}\vec q_\perp\cdot\vec b}
-e^{-i{1\over 2}\vec q_\perp\cdot\vec b}\right).
\label{eq:chi1}
\end{equation}
This expression can be obtained for any ratio of quark masses by using
different shifts of $\vec B$. An examination of Eq.~(\ref{eq:chi1})
shows that $\chi(B,b,\hat B\cdot\hat b)$ is an odd function of
$\hat B\cdot\hat b\equiv cos(\phi)$.
The real part of $\hat f(\theta=0,b)$ is
proportional to the integral of $sin(\chi(B,b,\hat B\cdot\hat b))$
over $\phi$ between 0 and $2\pi$ and therefore vanishes.
Thus we have a general theorem that the forward scattering amplitude
for the scattering of two neutral systems (made of two particles)
that interact via Coulomb forces
is purely imaginary.

We now evaluate the imaginary part of $\hat f(\theta=0,b)$.
A closed form expression for the function
$\chi(B,b,\hat B\cdot\hat b)$ can be obtained by performing the integral
over $d^2q_\perp$ in Eq.(\ref{eq:chi1}). The result is
\begin{eqnarray}
\chi(B,b,\hat B\cdot\hat b)= 2\alpha/v[K_0(3x_2)-K_0(6x_2)
+K_0(6x_1)-K_0(3x_1)+                           \nonumber\\
{1\over 2}(3x_2K_1(3x_2)-6x_2K_1(6x_2)+
6x_1K_1(6x_1)-3x_1K_1(3x_1))],
\end{eqnarray}
where
$x_1=\mid \vec B+{1\over 2}\vec b\mid/r_B$,
$x_2=\mid \vec B-{1\over 2}\vec b\mid/r_B$ and $K_i$ are modified Bessel
functions. This expression can be used to evaluate the imaginary part of
$\hat f$. The results are shown in the  Fig. 1.

We see that one finds a $b^2$ behavior for small values of b,
and in that regime the amplitude is proportional to $\alpha^2$. This
indicates that perturbation theory is valid at small b, even though the
coupling constant $\alpha$ can be large.

Our purpose in this paper is to focus on the properties of wave packets of
small
transverse size. In particular, Eq.~(\ref{eq:sigma_theory}) shows that for
large enough Q, the effective value of $b^2\sim 1/Q^2$, which is small.
Thus we use a simplified version of the interaction
\begin{equation}
\hat \chi(b)\equiv \hat f(\theta=0)\approx \sigma b^2/<b^2>,
\end{equation}
where $<b^2>$ represents the ground state expectation value of the
operator $b^2$.

The calculations of the present section are a Coulomb version of the
calculations of Refs.~\cite{Low75,gs77,Nus75} which used one gluon exchange.
Those references also  found
a $b^2$ behavior for projectiles of small transverse
size.  The origin is color neutrality, which we have modeled here as electrical
neutrality. We have made non-perturbative calculations of all orders in
$\alpha$ which we have taken as large as three.

\section{CALCULATION}
The expression (\ref{eq:sigma}) for $\sigma_{eff}(Z)$ is evaluated
numerically. The matrix elements $\hat{b}^{2}_{X}$ and form factors
$f_{X}(Q^{2})$ are calculated for 600 discrete
states and a $[0,10]$ range of the continuous variable $k$. This is
shown to be sufficient to reproduce the analytic result,
Eq.~(\ref{eq:sigma_theory}) for momentum transfer $Q r_B$ of up to 20.
These calculations are performed for $P=Q$ which corresponds to the
quasielastic kinematics of Ref.~\cite{ne18}. 
The results for the real part of $\sigma_{eff}(Z)$ for several values
of the momentum transfer $Q$ and $\alpha=2$ are shown in Fig. 2.
The ``nucleon" expands more slowly (is large for a larger value of Z)
 for higher values of the
momentum transfer. There is an initial drop, which is analysed below and
due primarily to the contribution of the states with $l=2$.
After the initial drop for a wide range of Z
 the effective cross section $\sim Z$, which is consistent with the
``linear expansion" model
of Refs. \cite{FS88,FLFS89} and also the results of $\cite{JM1}$.

The Z-dependence for a given value of $Q r_B=6$  and three values of
$\alpha=1/137,2,3$,   is presented in Fig. 3.
The figure shows that the
``nucleon"  expansion distance grows with $\alpha$.
This is because $P_X$ moves closer to $P$ as $\alpha$ increases, recall
eqs.~(\ref{eq:P_X}).  
 This is discussed below in more detail.

To quantitatively describe the expansion let's introduce an expansion
distance $Z_{exp}$  defined as follows
\begin{equation}
\sigma_{eff}(Z_{exp}) = \sigma.
\end{equation}
The expansion distance $Z_{exp}$  is
shown as a function of the momentum transfer
$Q$ in Fig.4. We
 see that $Z_{exp}$ is linear with $Q$; the value of $\alpha$ determines the
slope.

We calculate
$\sigma_{eff}(Z)$
for a reduced range of the continuous variable $k\in[0,k_{max}]$
to further investigate the importance of the higher excited states.
If $k_{max}$ is such that $P(k_{max})=k_{max}\sim\sqrt{Q}$ then the reduced
$\sigma_{eff}(Z)$ is almost identical to the full one. This result is
important to support the validity of the eikonal approximation.
 The $\sigma_{eff}(Z)$ for $k_{max}=1$ is
shown in Fig.5. Even though on the initial stage the expansion picture is
different, for the most of the range it reproduces the expansion of the
``full" PLC well and the expansion
distance is nearly the same as for the ``full" PLC.

\section{DISCUSSION}

A conclusion can be made that a transparency phenomenon
is obtained for the model under
consideration. A small size object is initially formed in a high momentum
transfer process and the expansion rate of this object is inversely
proportional to the momentum transfer $Q$.
While the first result has been known for this model
for some time \cite{FMS92,FMS931}, the second result is a  new consequence
for this model.
We therefore take a
closer look at what causes such a decrease (favorable for color transparency)
in the
expansion rate.
As was mentioned above, PLC expansion
depends on the momenta of the intermediate
states $P_X$.
For $\Delta M_X/\alpha \ll Q$   (~\ref{eq:P_X}) can be expanded as
\begin{equation}
P_X= Q+ \frac{\Delta M_X/\alpha}{2Q} +...
\end{equation}
Thus  the values of $\Delta M_X$, recall Eq.~(\ref{eq:P_X}),
that correspond to the intermediate states X which make important
contributions to the sums in Eq.~(\ref{eq:sigma})
determine the expansion rate.
These states have a discrete or continuum nature.
But the energies of the exited discrete states
are bounded from above and their contributions to
$\sigma_{eff}(Z)$; $b^2_{nl}$ and $f_{nl}(Q^2)$ decay rapidly and
monotonically with the number $n$. Thus it is more relevant to examine the
contributions of the continuum states, which in principle can have very high
energies.

The form factors for the continuum
$f_{kl}(Q^2)$ display a peaking behavior, which occurs when the momentum
transfer Q matches the relative momentum denoted by the quantum number k.
See Fig.6 which shows also that
the energy of the states produced in a high momentum
transfer process grows linearly with $Q$. If these states had appreciable
matrix elements $b^2_{kl}$ for the soft interaction, the expansion would be
very rapid.
However,  $b^2_{kl}$ decays rapidly with
 $k$ (Fig.7). The $b_{k2}^2$ and $f_{k2}$ matrix elements
exhibit similar behaviour.
As a result $b^2_{kl} f_{kl}(Q^2)$ have maxima that experience only slight
increase with $Q$ (Fig.8).
 So we see that in this model high excited states are formed for high
momentum transfer, but the soft interaction cuts them off.

The $k-$dependence of $b_{kl}^2 f_{kl}$ determines the expansion of the
PLC. We have seen in Figs. 3 and 5 that there is an initial drop
for small values of $Z$.  
We argue that this drop is caused by the contribution of the
states with $l=2$.
To see this we expand Eq.~(\ref{eq:sigma}) for small $Z$:
\begin{eqnarray}
Re(\sigma_{eff}(Z))= \sigma
+ \sum_{l=0,2} \int_{k_{thr}}^{\infty} dk\frac{\chi_{kl}
f_{kl}(Q^{2})}{F(Q^{2})}-          \nonumber\\
Z \sum_{l=0,2} \int_{k_{thr}}^{\infty} dk\frac{\chi_{kl}
f_{kl}(Q^{2})}{F(Q^{2})}ImP_{k}-          \nonumber\\
Z^2 \sum_{l=0,2} \int_{0}^{\infty} dk\frac{\chi_{kl}
  f_{kl}(Q^{2})}{F(Q^{2})}Re(P_{k}-P)^2
\label{eq:drop}
\end{eqnarray}
Here we omit, for simplicity, the contribution of the discrete states;
$k_{thr}$ is defined by $P_{k_{thr}}=0$. The calculations show that
$b_{k0}^2 f_{k0}<0$, whereas $b_{k2}^2 f_{k2}>0$. In the first integral in
(~\ref{eq:drop}) the states with $l=0$ dominate, which results in the small
 $\sigma_{eff}(0)$. In the second and third integral the situation
is reversed, although with the similar result:
the dominating $l=2$ states cause the initial
shrinkage of the PLC.

Another piece of information is provided by the $\alpha$ dependence of the
expansion distance. The latter decreases with the
$\Delta M_{X}$ of the important states $|X>$. For all the discrete states
and for the continuum states with $k<1$ $\Delta M_{X}$
decrease with
$\alpha$. Since $Z_{exp}$ grows with $\alpha$, it can be seen
that only the states with $k<1$ are important for the PLC expansion.

This analysis leads to the conclusion that the maximum energy of the states
 which are relevant for the PLC expansion grows much slower than
the momentum transfer $Q$. This result is important in two ways. First, it
illustrates that this indeed is the condition necessary for CT to occur.
Second, our approach was based on the eikonal approximation, which assumed
the momenta of the propagating states to be much greater than the size of
the system. This approximation breaks down for the states with energy close
to the momentum transfer $Q$. But since these states are not relevant and
their contribution is negligibly small, the validity of the
eikonal approximation is proven by the above conclusion.

  \section{CONCLUSION}

 A ``nucleon"  model of a hadron has been investigated. In this
model hadron consists of  two quarks bound by the Coulomb potential with
the variable strength.
A small transverse size object is formed when ``nucleon" absorbs
a high momentum  photon.
Such a system expands with the rate inversely proportional to the
momentum transfer $Q$.
This slow expansion is a consequence of the fact
 that the states with
energy much greater than the ground state energy are not very important.

\newpage
\noindent{\bf Figure Captions}

\begin{itemize}

\item Figure 1. Imaginary part of the forward scattering amplitude $\hat f$ for
three
different values of $\alpha/v$.

\item Figure 2. PLC expansion for 4 different momenta Q. $\alpha=2$.

\item Figure 3. PLC expansion for three values of $\alpha$. Q = 6.

\item Figure 4. Expansion distance $Z_{exp}$ for three values of $\alpha$.

\item Figure 5. $\sigma_{eff}(Z)$ for the ``full" PLC and a ``reduced" PLC.
Full PLC is constructed out of 600 discrete states and [0,10] range
of the continuous spectrum.  Reduced PLC is constructed out of 600
discrete states and [0,1] range of the continuous spectrum.  $\alpha$ = 2,
Q = 6.

\item Figure 6.Inelastic form factors $f_{k0}(Q^2)$ for ten values of $Q$.

\item Figure 7. $b^2_{k0}$ decays rapidly with $k$.

\item Figure 8. $b^2_{k0}f_{k0}$ for the same ten values of $Q$ as in
Fig. 6.

\end{itemize}
\end{document}